\documentclass[letterpaper, 10pt, conference]{ieeeconf}  

\usepackage[T1]{fontenc}
\usepackage{CJKutf8}
\usepackage[english]{babel}
\usepackage{tipa}
\usepackage{multirow}
\usepackage{setspace}
\usepackage{comment}
\usepackage{cite}
\usepackage{graphicx}
\usepackage{booktabs}
\usepackage{mathtools}
\usepackage{url}

\IEEEoverridecommandlockouts                              

\title{\LARGE \bf
BrainTalker: Low-Resource Brain-to-Speech Synthesis \\with Transfer Learning using Wav2Vec 2.0}
\author{Miseul Kim, Zhenyu Piao, Jihyun Lee and Hong-Goo Kang$^{\dagger}$
\thanks{This work was supported by the project ‘Alchemist Brain to X (B2X) Project’ through the Ministry of Trade, Industry and Energy (MOTIE), South Korea, under Grant 20012355 and NTIS 1415181023.}%
\thanks{All authors are with Digital Signal Processing \& Artificial Intelligence Lab, School of Electrical and Electronic Engineering, Yonsei University, Seoul, South Korea {\tt\small \{miseul4345, jkyung, leeji0526\}@dsp.yonsei.ac.kr, hgkang@yonsei.ac.kr}}
\thanks{$\dagger$Corresponding author}
}

\begin{document}

\maketitle
\thispagestyle{empty}
\pagestyle{empty}

\begin{abstract}


Decoding spoken speech from neural activity in the brain is a fast-emerging research topic, as it could enable communication for people who have difficulties with producing audible speech.
For this task, electrocorticography (ECoG) is a common method for recording brain activity with high temporal resolution and high spatial precision.
However, due to the risky surgical procedure required for obtaining ECoG recordings, relatively little of this data has been collected, and the amount is insufficient to train a neural network-based Brain-to-Speech (BTS) system.
To address this problem, we propose BrainTalker---a novel BTS framework that generates intelligible spoken speech from ECoG signals under extremely low-resource scenarios.
We apply a transfer learning approach utilizing a pre-trained self-supervised model, Wav2Vec 2.0.
Specifically, we train an encoder module to map ECoG signals to latent embeddings that match Wav2Vec 2.0 representations of the corresponding spoken speech.
These embeddings are then transformed into mel-spectrograms using stacked convolutional and transformer-based layers, which are fed into a neural vocoder to synthesize speech waveform.
Experimental results demonstrate our proposed framework achieves outstanding performance in terms of subjective and objective metrics, including a Pearson correlation coefficient of 0.9 between generated and ground truth mel-spectrograms.
We share publicly available Demos and Code~\footnote{\url{https://braintalker.github.io/}}.


\end{abstract}

\section{Introduction}
\label{lab:intro}

\begin{figure*}[t]
\centering
\centerline{\includegraphics[width=16.5cm]{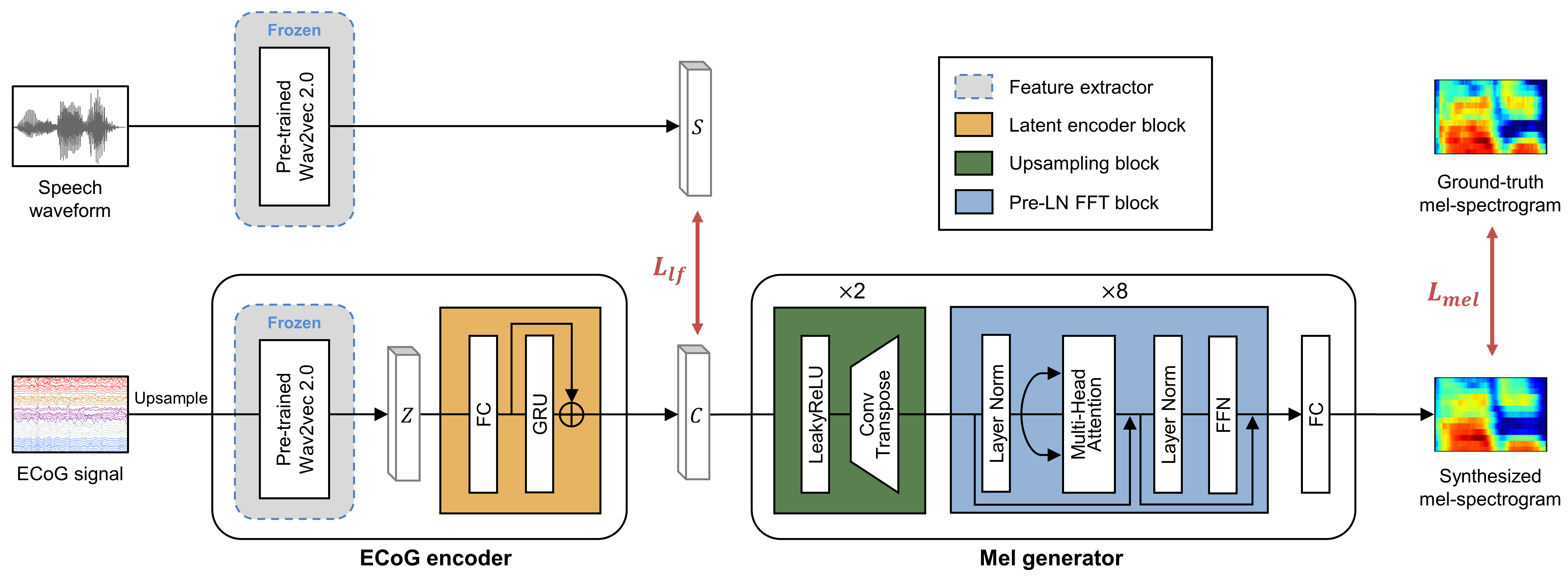}}
\caption{Overall structure of the proposed framework. A coarse brain representation $Z$ is extracted from a pre-trained Wav2Vec 2.0 feature extractor. Then, a latent encoder block generates a new feature $C$. $C$ is passed through upsampling and FFT blocks in order to generate a mel-spectrogram, which is then vocoded to produce speech.}  
\label{fig:proposed}
\end{figure*}

Brain-to-speech (BTS) technology involves the use of a computer system to translate neural activity in the brain into words or sentences~\cite{herff2015brain, wilson2020decoding, lee2023towards}.
Decoding spoken speech from the brain is an especially desirable technology for individuals with disabilities or conditions that affect their ability to speak.
However, the intricate nature of brain signals poses significant challenges to analyzing and extracting speech-related information from neural activity~\cite{furman2019modern, gao2021complex, kumar2012analysis}.
As such, research on decoding spoken speech from the brain is still in its early stages in terms of signal representation choices, architectural design, and decoded speech quality.

Electrocorticography (ECoG)~\cite{dubey2019cortical} recordings have emerged as a promising candidate for brain activity representations in BTS research.
ECoG signals are typically obtained through invasive recording methods such as implanting subdural grid electrodes~\cite{dubey2019cortical, buzsaki2012origin}.
Therefore, they have higher spatio-temporal resolution and noise robustness compared to non-invasive recording methods such as electroencephalography (EEG)~\cite{sanei2013eeg}.

There have been several prior approaches utilizing ECoG signals to decode spoken speech from the brain.
\cite{anumanchipalli2019speech} took a two-stage decoding approach based on long short-term memory (bi-LSTM)~\cite{graves2012long} in which articulatory kinematic features are estimated from the ECoG signals, and mel-frequency cepstral coefficients (MFCCs) are decoded from the estimated articulatory features.
However, the system needed to be trained on large amounts of brain and speech signals.
In practice, it is difficult to collect a large amount of ECoG data from many patients since the recording process is invasive, as well as physically and mentally demanding for participants.
Given these low-resource constraints, other works have adopted existing high-performance architectures to synthesize intelligible speech.
\cite{angrick2019speech} employed a Densely-Connected Convolutional Network (DenseNet) \cite{huang2017densely} to generate mel-spectrograms from ECoG signals and used a Wavenet vocoder~\cite{DBLP:conf/ssw/OordDZSVGKSK16} to synthesize speech waveforms.
\cite{10097004} adopted a combination of convolutional layers, transformer~\cite{vaswani2017attention}, and a ParallelWaveGAN vocoder~\cite{yamamoto2020parallel} for the same purpose.
Although both works were able to successfully synthesize speech signals with only a few minutes of ECoG recordings, the quality of the synthesized speech signals was unsatisfactory.
For example, the two works reported the maximum Pearson correlation coefficients between an estimated mel-spectrogram and a ground-truth mel-spectrogram as 0.69 and 0.75, respectively.
As such, only a few attempts have successfully synthesized spoken speech using low-resource ECoG data.

In this work, we propose a novel framework to effectively address the data shortage issue and synthesize high-quality speech from ECoG signals, which we call \textbf{BrainTalker}.
We adopt a transfer learning strategy in which we use a pre-trained self-supervised speech representation model, Wav2Vec 2.0~\cite{baevski2020wav2vec}, as part of an ECoG encoder.
The purpose of this encoder is to extract coarse brain representations that contain generalized information from the ECoG data.
This is done by incorporating a latent feature loss, which induces the ECoG encoder to produce representations that match the Wav2Vec 2.0 representations of the corresponding spoken speech.
Then, the brain representations are passed through a generator network that produces mel-spectrograms, which are in turn fed into a pre-trained HiFi-GAN vocoder~\cite{kong2020HiFi} to generate speech waveforms.
Using this framework, our proposed model is able to generate speech from brain signals even with only extremely small amounts of ECoG data for both seen and unseen word generations. 

In summary, the contributions of this paper are as follows:
\begin{itemize}
    \item \textbf{Novel framework:} 
    We propose a novel BTS architecture by adopting transfer learning from Wav2Vec 2.0. 
    To the best of our knowledge, this is the first attempt to utilize transfer learning from a pre-trained speech representation model to obtain brain representations from ECoG signals for spoken speech synthesis, and it allows us to effectively address the data shortage problem.
    \vspace{2pt}
    
    \item{\textbf{New training strategy:}} \
    To extract speech-related informative representations from brain signals, we introduce a new training criterion called latent feature loss. This criterion encourages embeddings extracted from brain signals to closely match the corresponding spoken speech representations.   
    \vspace{2pt}

    \item \textbf{Performance:} Our proposed framework enables the decoding of spoken speech that is highly correlated with ground-truth mel-spectrograms for both seen and unseen ECoG data. 
\end{itemize}

\vspace{-8pt}
\section{Proposed model}
\label{sec:proposed_model}
In this section, we provide a detailed explanation of our proposed BrainTalker architecture. 
As shown in Figure~\ref{fig:proposed}, the overall framework mainly consists of two modules: an ECoG encoder and a mel generator.
The ECoG encoder consists of a pre-trained Wav2Vec 2.0 feature extractor and a latent encoder block. 
The Wav2Vec 2.0 model is used to extract coarse brain features $Z$ from the ECoG signals.
The latent encoder block uses $Z$ to produce new representations $C$.
The representations $C$ are forced to be similar to speech representations $S$, which are obtained by passing speech waveforms corresponding to the ECoG signals through Wav2Vec 2.0.
The ECoG generator is trained to closely match $C$ to $S$ using our proposed latent feature loss (Section \ref{subsec:criteria}).
Then, a mel generator network uses $C$ to produce mel-spectrograms.
The mel generator consists of upsampling and pre-layer normalization feed-forward Transformer (pre-LN FFT) blocks.
Finally, the estimated mel-spectrograms are fed into a pre-trained vocoder to produce raw speech waveforms. 

\subsection{ECoG encoder}
\noindent\textbf{Feature extractor.}
Wav2Vec 2.0 is a speech representation model that is pre-trained in a self-supervised manner~\cite{baevski2020wav2vec}.
Although it was originally trained on and designed for speech data, Wav2Vec 2.0 representations contain an abundance of information that is generally valuable for modeling sequential data~\cite{Ou2022TowardsTL, kostas2021bendr}. 
Inspired by this observation, we adopt Wav2Vec 2.0 as a feature extractor to obtain coarse brain representations from ECoG recordings.

To use ECoG signals as the inputs to a pre-trained Wav2Vec 2.0 model, the sampling rate of the ECoG signals has to be consistent with that of the model's original training data. 
Therefore, we first upsample ECoG signals from 2kHz to 16kHz.
Since ECoG signals consist of multiple channels depending on the number of recording sensors, each channel is passed through the feature extractor independently and all outputs are stacked across the channel dimension to form the coarse brain representations $Z$.
\vspace{2pt}

\noindent\textbf{Latent encoder block.}
The latent encoder block aims to transform the coarse brain features $Z$ into new representations $C$ that can be used for generating mel-spectrograms.
To integrate the information from all ECoG channels, we employ a fully connected (FC) layer to map all the channels of $Z$ into one. 
Then, a gated recurrent unit (GRU) is used to extract global information from $Z$, along with a residual connection to avoid the vanishing gradient problem during training.

\subsection{Mel generator}
The objective of the mel generator is to estimate mel-spectrograms given the representations $C$. 
The mel generator consists of the following three submodules:

\vspace{2pt}
\begin{enumerate}
    \item \textbf{Upsampling block.}
    We upsample $C$ using 2 transposed convolutional layers to match the temporal resolution of $C$ to that of the output mel-spectrograms. 
    To provide a non-linearity to the block, Leaky ReLU~\cite{xu2020reluplex} activation is added before the transposed convolution.

    \item \textbf{Pre-LN FFT block.}
    We apply 8 feed-forward Transformer (FFT) blocks~\cite{vaswani2017attention} together with the upsampling blocks to encode the long-term dependencies from the hidden features.
    A conventional FFT block consists of two modules: multi-head self-attention layers and a feed-forward layer.
    Each module is followed by layer normalization~\cite{xiong2020layer} and a residual connection~\cite{he2016deep}. 
    Unlike conventional FFT blocks, we adopt pre-LN FFT blocks, that place the layer normalization inside the residual blocks instead of between them~\cite{xiong2020layer}; this has been proved to improve the stability of training \cite{10095477}.
    \vspace{2pt}

    \item \textbf{FC layer.}
    Finally, the outputs from the FFT blocks are passed through a fully connected layer to map the hidden dimension of the features to those of the target mel-spectrograms.
\end{enumerate}

\subsection{Training criteria}
\label{subsec:criteria}
We use two criteria for the overall training process: mel loss ($L_{mel}$) and latent feature loss ($L_{lf}$).
The latent encoder block and mel generator are simultaneously trained using both $L_{mel}$ and $L_{lf}$.
Note that parameters of the Wav2Vec 2.0 feature extractors are frozen during the entire training procedure.
\vspace{2pt}

\noindent\textbf{Mel loss.}
We minimize the $L_{2}$ distance between the generated mel-spectrogram $\tilde{X}$ and the target mel-spectrogram $X$. $L_{mel}$ is computed as follows: 
\begin{equation}
L_{mel} = \lVert \tilde{X} - X \rVert_{2}.
\end{equation}

\vspace{2pt}
\noindent\textbf{Latent feature loss.}
The main idea behind our framework is to extract speech-related information from brain signals.
To do this, we aim to minimize the information gap between the brain representations $C$ and corresponding speech representations $S$ using a novel latent feature loss $L_{lf}$.
$L_{lf}$ reduces the $L_{2}$ distance between $C$ and $S$ so as to make $C$ contain suitable information for mel-spectrogram generation; $L_{lf}$ is calculated as:
\begin{equation}
L_{lf} = \lVert C - S \rVert_{2}.
\end{equation}
Therefore, the overall training loss is:
\begin{equation}
L_{tot} = \lambda_{mel}L_{mel} + \lambda_{lf}L_{lf}.
\end{equation}
where $\lambda_{mel}$ and $\lambda_{lf}$ are weights of losses and set as 1 in experiments.

\section{Experiments}
\label{sec:experiment}


\subsection{Data recording protocol}
Data recording was conducted with a 55-year-old male subject who had undergone a neurosurgical procedure for epilepsy.
Informed consent was obtained from the participant, and the entire recording process complied with all relevant ethical regulations.
In data collection, the subject listened to a series of simple questions and was asked to verbally respond to the questions by selecting one of two given options.
All recording steps were done in Korean.
ECoG and speech signals were recorded simultaneously at sampling rates of 2kHz and 16kHz, respectively.

There were 12 different words in the answering options: \textipa{[kaŋadzi], [s\textsuperscript{h}aram], [kom], [nuns'aram], [modza], [ts\textsuperscript{h}\textepsilon k\textcorner], [nun], [k\textsuperscript{h}o], [s\textsuperscript{h}on], [k'ot\textcorner], [k\textsuperscript{h}\textturnv p\textcorner], [ts\textsuperscript{h}oŋ]}.
During the recording process, the subject repeated each word 12 times in total.
We set aside all repetitions of the word \textipa{[s\textsuperscript{h}on]} for unseen word generation because all of the phonemes in this word appeared in the other words.
We additionally excluded a single trial of the 11 seen words for seen data generation.
After eliminating a few mispronounced words spoken by the participant, the total amount of training data was about 1 minute, which is an extremely small amount for a speech synthesis task.

\begin{figure}[t]
\centering
\centerline{\includegraphics[width=8.7cm]{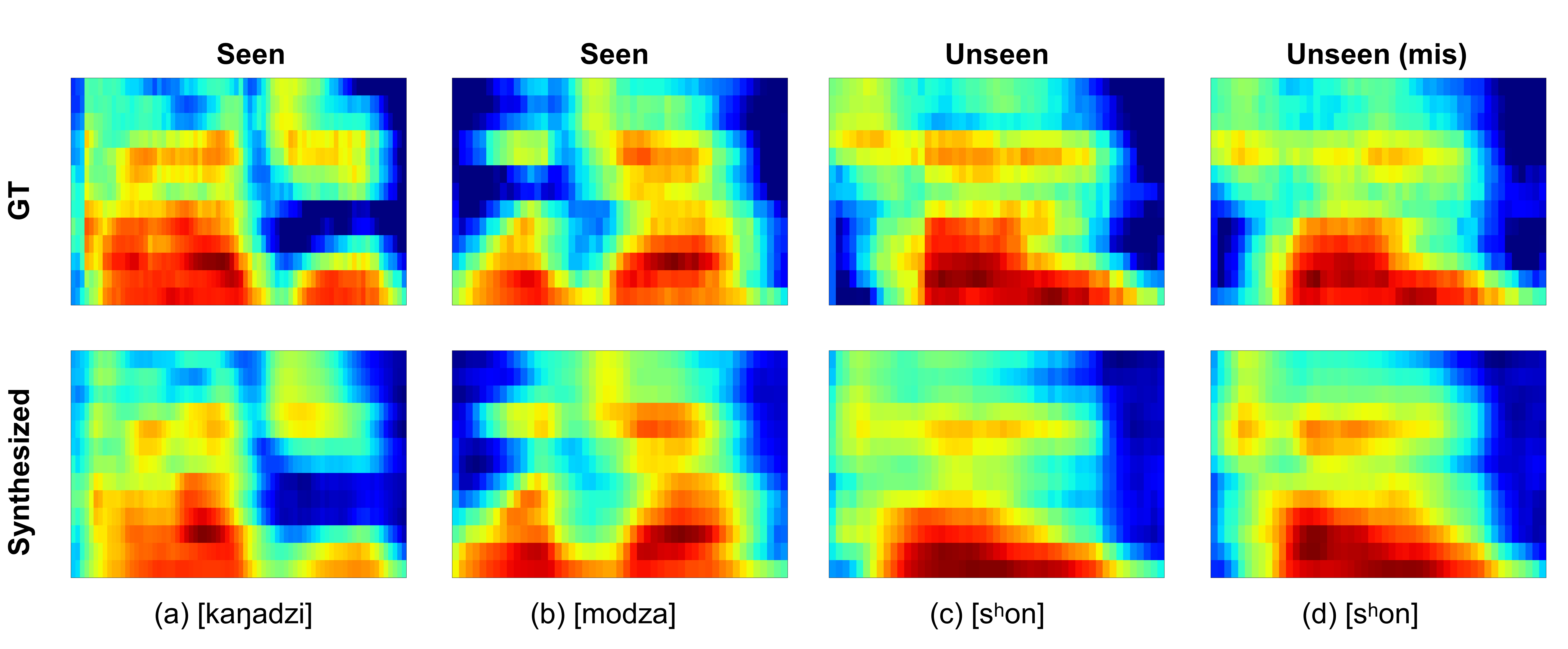}}
\caption{Ground-truth (GT) and synthesized mel-spectrograms corresponding to seen and unseen words produced by our proposed system. Figures (a), (b), and (c) show successfully synthesized samples. (d) is an incorrectly synthesized sample, which mis-pronounces \textipa{[s\textsuperscript{h}on]} as \textipa{[ts\textsuperscript{h}oŋ]}.}
\label{fig:spectrogram}
\end{figure}
\vspace{-6pt}

\subsection{Implementation details}
We used ECoG signals recorded from the Parietal cortex area in the left hemisphere, which is known to be closely related to speech production\cite{rauschecker2009maps, shum2011sensorimotor, geranmayeh2012contribution}.
An 8-contact strip electrode was attached to record neural activity from this area.
We set 13-dimensional mel-spectrograms as the target of the mel generator.
The dimension of the feature is relatively low, but we chose it because of difficulties with predicting higher-resolution mel-spectrograms given the limited data available.
Mel-spectrograms were obtained using a 25 ms window length and a 10 ms hop length.
The model was trained using the Adam optimizer \cite{Adam} with an initial learning rate of 0.00005, and the learning rate was scheduled to decay by multiplying 0.9 after every 100 epochs with the StepLR scheduler.
We adopted HiFi-GAN~\cite{kong2020HiFi} as a neural vocoder to generate speech waveforms from the estimated mel-spectrograms.
The vocoder was pre-trained using the VCTK Corpus~\cite{veaux2017cstr} and public Korean datasets.\footnote{https://aihub.or.kr/aihubdata/data/view.do?\\currMenu=115\&topMenu=100\&aihubDataSe=realm\&dataSetSn=109}

For a baseline model, we adopted a recently published ECoG-to-speech synthesis architecture~\cite{10097004}.
The baseline uses one temporal convolution layer and 8 transformer encoder layers to decode 80-dimensional mel-spectrograms from ECoG signals.
Parallel WaveGAN~\cite{yamamoto2020parallel} is then used to generate raw speech waveforms from estimated mel-spectrograms.
To ensure a fair comparison, we modified the preprocessing of the input waveform and target mel-spectrograms in the baseline to correspond with the processing in the proposed model.
We also used HiFi-GAN as the vocoder for all baseline experiments.

\subsection{Evaluation metrics}
We evaluated the synthesis performance of the proposed model using both subjective and objective metrics. 
For subjective evaluations of speech naturalness, we performed mean opinion score (MOS) tests~\cite{streijl2016mean}.
A total of 19 participants were asked to provide discrete scores on a scale from 1 to 5. 
Each respondent evaluated a total of 22 test words, which consisted of 11 samples of seen words and 11 samples of the unseen word (pronouncing \textipa{[s\textsuperscript{h}on]}).

For objective evaluations of speech quality, we measured root mean-squared error (RMSE), mel cepstral distance (MCD)~\cite{kubichek1993mel}, and Pearson correlation coefficient (PCC) between generated and ground-truth mel-spectrograms.
RMSE and MCD measure absolute and perceptually oriented accuracies of the predicted mel-spectrograms, respectively.
PCC measures how closely the predicted spectrogram is correlated with the actual spectrogram.
We report all values in the tables with 95\% confidence intervals.

\begin{table}[t]
\centering
\caption{Speech decoding performance in terms of subjective metrics. GT: ground-truth speech. GT (HiFi-GAN): samples generated using pre-trained HiFi-GAN vocoder with ground-truth mel-spectrograms. Proposed: synthesized samples from the proposed model.}
\resizebox{\columnwidth}{!}
{%
\begin{tabular}{cccc}
\toprule
Model        & Seen & Unseen & Total \\ \midrule
GT           & -    & -    &3.88±0.10    \\
GT (HiFi-GAN) & -    & -   &3.04±0.09    \\ \midrule
Baseline~\cite{10097004}   &1.81±0.12   &1.87±0.12     &1.84±0.08    \\
BrainTalker (Proposed)                   &\textbf{2.31±0.13}    &\textbf{2.32±0.13}     &\textbf{2.31±0.09}     \\ \bottomrule

\end{tabular}%
}
\label{tab:tab_1}
\end{table}

\subsection{Experimental results}
\noindent\textbf{Subjective quality.}
Table~\ref{tab:tab_1} shows speech synthesis performance assessed on subjective metrics for samples from the ground truth (GT), baseline, and proposed model.
We also include reconstruction results from GT using the HiFi-GAN vocoder as an additional reference point.
BrainTalker obtains a total MOS of 2.31, significantly outperforming the baseline that achieves 1.84.
In addition, it obtains almost identical performance for both seen and unseen word generation (2.31 and 2.32, respectively).
These results imply that, when the model is applied to phonemes that were seen during training, it can effectively synthesize unseen words as well.
We note that there is a ceiling to our model's performance because the speech reconstruction performance of the HiFi-GAN vocoder is not perfect (3.88 MOS for GT vs. 3.04 for GT (HiFi-GAN)).
This is due to using a low-dimensional mel-spectrogram as the input feature of the vocoder.
\vspace{2pt}

\noindent\textbf{Objective quality.}
Table~\ref{tab:tab_2} shows the performance of BrainTalker compared against the baseline using the RMSE, MCD, and PCC objective metrics.
The results show that our proposed model outperforms the baseline in terms of all three metrics for synthesizing the 13-dimensional mel-spectrograms.
Notably, our model achieves exceptional performance in terms of PCC, obtaining a score of 0.9.

Also, to verify the validity of using 13-dimensional mel-spectrograms rather than higher-dimensional ones in the extremely low-resource scenario, we conducted additional experiments with a mel generator trained to estimate 80-dimensional mel-spectrograms. 
This approach aligned with the original framework employed in the baseline~\cite{10097004}.
From the results, we observe that for both the baseline and the proposed architectures, using 13-dimensional mel-spectrograms leads to higher performance in terms of RMSE and PCC.
Thus, we conclude that the models could not effectively reconstruct the high-frequency components of 80-dimensional mel-spectrograms using the low-resource data and result in large errors between the estimated and ground-truth vocoding features.

\begin{table}[t]
\centering
\caption{Speech decoding performance of BrainTalker compared with the baseline in terms of objective metrics. The number next to the model name represents the number of mel-bin. 
} 
\resizebox{\columnwidth}{!}{%
\begin{tabular}{cccc}
\toprule
Model                & RMSE(↓) & MCD(↓)        & PCC(↑)   \\ \midrule
Baseline-13          & 1.69±0.17    & 6.63±0.75    & 0.86±0.04          \\ 
Baseline-80          & 1.73±0.17    & 6.40±0.95    & 0.66±0.12          \\ \midrule
\textbf{\textit{BrainTalker-13 (Proposed)}}       & \textbf{1.28±0.16}    & 5.64±0.88    & \textbf{0.90±0.04}          \\
\textbf{\textit{BrainTalker-80}}      & 1.40±0.16    & \textbf{5.32±0.70} & 0.72±0.08 \\ 
w/o Wav2Vec 2.0    & 1.36±0.17     & 6.43±0.86            & 0.86±0.06            \\ 
w/o $L_{lf}$   & 1.29±0.12     & 5.77±0.64            & 0.88±0.04            \\ \bottomrule
\end{tabular}%
}
\label{tab:tab_2}
\end{table}
\vspace{-4pt}

\vspace{2pt}
\noindent\textbf{Ablation studies.}
To validate the effectiveness of each component of the proposed model, we conducted ablation studies on certain components; the results are shown in the last two rows of Table~\ref{tab:tab_2}.
First, in order to confirm the effectiveness of our transfer learning strategy, we replaced the pre-trained Wav2Vec 2.0 feature extractor with a separate model with newly initialized weights.
This feature extractor contains 12 FFT blocks with an embedding dimension of 512 and a hidden dimension of 128, which is a much smaller network compared to Wav2Vec 2.0 (embedding dimension of 768 and hidden dimension of 3,072).
Then, we trained the entire framework from scratch. 
Replacing the pre-trained feature extractor leads to performance degradation in terms of all metrics compared to our proposed model.
Second, we trained our model without the latent feature loss $L_{lf}$ in order to validate its effectiveness.
We find that removing $L_{lf}$ also leads to a performance decrease across all of the objective metrics.

\vspace{2pt}
\noindent\textbf{Synthesized spectrogram analysis.}
In Figure~\ref{fig:spectrogram}, we show ground-truth mel-spectrograms along with ones predicted by our model, including both seen and unseen words. 
In most cases, it can be observed that our model is effective at generating high-quality 13-dimensional mel-spectrograms from the ECoG signal for both seen and unseen words (Figure~\ref{fig:spectrogram}-(a), (b), (c)). 
However, in certain unseen-word instances, our model occasionally generates partially inaccurate mel-spectrograms.
For example, in Figure~\ref{fig:spectrogram}-(d), synthesized mel-spectrogram represents a slightly altered pronunciation (\textipa{[ts\textsuperscript{h}oŋ]}) from an intended word (\textipa{[s\textsuperscript{h}on]}).
Despite these marginal errors, we can observe that BrainTalker still correctly captures some phonemes contained in the original words such as the \textipa{[s\textsuperscript{h}]} and \textipa{[o]}.

\section{Conclusion}
\label{sec:conclusion}

In this paper, we proposed a novel brain-to-speech framework called BrainTalker, which can generate spoken speech from ECoG data under extremely low-resource scenarios.
To effectively utilize the information in brain signals, we adopted the pre-trained self-supervised speech representation model Wav2Vec 2.0 to extract coarse brain latent representations. 
We proposed a new training criterion, latent feature loss, to guide the brain representations to be similar to Wav2Vec 2.0 representations of the raw speech corresponding to the ECoG signals.
Our model successfully synthesizes audible, intelligible speech from ECoG signals for words that were both seen and unseen during training, outperforming a recently proposed baseline in terms of both subjective and objective measurements.
Potential directions for future work include methods for improving the perceptual quality of synthesized speech, as well as expanding to a multi-speaker scenario. 



\bibliographystyle{IEEEtran}
\bibliography{my_ref}

\end{document}